\documentclass[preprint, 12pt]{aastex}

\usepackage{emulateapj5}
\usepackage{psfig}

\def\arcsec{$\,^{\prime\prime}$~}

\def\erg/cm2sec{ergs~cm$^{-2}$~s$^{-1}$}  
\def\ergcm2{ergs~cm$^{-2}$}

\newcommand{\lsim }{{\lower0.8ex\hbox{$\buildrel <\over\sim$}}}
\newcommand{\gsim }{{\lower0.8ex\hbox{$\buildrel >\over\sim$}}}

\def\apj{ ApJ}
\def\aap{ A\&A}

\def\aj{AJ}
\def\apjs{ApJ Supp}

\def\Chandra{${\it Chandra}$\ }
\def\HST{${\it HST}$\ }

\def\simge{\mathrel{%
   \rlap{\raise 0.511ex \hbox{$>$}}{\lower 0.511ex \hbox{$\sim$}}}}
\def\simle{\mathrel{
   \rlap{\raise 0.511ex \hbox{$<$}}{\lower 0.511ex \hbox{$\sim$}}}}

\newcommand{\Msun}{\ifmmode {M_{\odot}}\else${M_{\odot}}$\fi}
\newcommand{\Lsun}{\ifmmode {L_{\odot}}\else${L_{\odot}}$\fi}
\newcommand{\Rsun}{\ifmmode {R_{\odot}}\else${R_{\odot}}$\fi}

\shorttitle{M80 and qLMXBs}
\shortauthors{Heinke et al.}

\begin{document}
\title{A \Chandra X-ray Study of the Globular Cluster M80}   

\author{C. O. Heinke,  J. E. Grindlay, P. D. Edmonds, D. A. Lloyd, 
S. S. Murray}
\affil{Harvard College Observatory,
60 Garden Street, Cambridge, MA  02138;\\
 cheinke@cfa.harvard.edu, josh@cfa.harvard.edu, pedmonds@cfa.harvard.edu,
  dlloyd@cfa.harvard.edu, ssm@cfa.harvard.edu}
\and
\author{H. N. Cohn and P. M. Lugger}
\affil{Department of Astronomy, Indiana University, Swain West 319,
Bloomington, IN 47405; cohn@indiana.edu, lugger@indiana.edu}

\slugcomment{Submitted to ApJ}

\begin{abstract}

We report our analysis of a \Chandra X-ray observation of the rich globular
cluster M80, in which we detect some 19 sources to a limiting 0.5-2.5
keV X-ray luminosity of $7\times10^{30}$ ergs s$^{-1}$ within the half-mass
radius.  X-ray spectra indicate that two of these sources are
quiescent low-mass X-ray binaries (qLMXBs) containing neutron stars. 
We identify five sources as probable cataclysmic variables (CVs), one of
which seems to be 
heavily absorbed, implying high inclination.  The brightest CV may be the 
X-ray counterpart of Nova 1860 T Sco.  The concentration of the
X-ray sources within the cluster core implies an average mass of
1.2$\pm0.2$ \Msun, consistent with 
the binary nature of these systems and very similar to the radial
distribution of the blue stragglers in this cluster.  The X-ray and
blue straggler source populations in M80 are compared to those in the
similar globular cluster 47 Tuc. 

\end{abstract}


\maketitle

\section{Introduction}

The \Chandra X-ray Observatory has allowed rapid gains in the study of
X-ray sources in globular clusters, especially when combined with the
resolution of the {\it Hubble Space Telescope} (\HST).  Faint X-ray
sources had been identified with {\it Einstein} (Hertz \& Grindlay
1983) and {\it ROSAT} (see Verbunt 2001 for a review).  A few of these
had been identified with bright LMXBs in quiescence (qLMXBs; e.g. Verbunt et
al. 1984) or with cataclysmic variables (Cool et al. 1995).  Recently
\Chandra (and to a lesser degree {\it XMM}) has allowed the
identification and detailed study of scores 
of faint X-ray sources in 47 Tuc
(Grindlay et al. 2001a, hereafter GHE01a), NGC 6397 (Grindlay et al. 2001b,
hereafter GHE01b),  $\omega$ Cen (Rutledge et al. 2002,
 Cool, Haggard, \& Carlin 2002),  NGC 6752 (Pooley et al. 2002b), NGC
6440 (Pooley et al. 2002a), and M28 (Becker et al. 
2003), among others.  Optical and radio identifications have allowed
secure identifications of cataclysmic variables (CVs),
chromospherically active binaries
(ABs), and millisecond pulsars (MSPs); see Edmonds et al. (2003a, b),
Grindlay et al. (2002) and Pooley et al. (2002b) for examples.  The
spectral and luminosity signatures of qLMXBs, thought to emit thermal
radiation from the neutron star surface (e.g. Brown, Bildsten \&
Rutledge 1998), allow them to be identified easily (e.g. Rutledge et
al. 2002, GHE01b).  These advances 
make it practical to compare significant populations of X-ray sources
in different globular clusters, exploring similarities or differences
in properties or formation mechanisms (see Pooley et al. 2003, Heinke
et al. 2003c). 

In this paper, we present new \Chandra observations of the globular 
cluster M80 (=NGC 6093).  This globular cluster has a small core
(6\farcs5, Ferraro et al. 1999) and relatively high
central density (log$(\rho_0)=4.87$, computed using the prescription
of Djorgovski 1993), although it is not core-collapsed.  The
distance to this cluster is estimated at 10.33$^{+0.8}_{-0.7}$ kpc
(Brocato et al. 1998), while $E(B-V)=0.17\pm0.01$, leading to a
neutral hydrogen 
 column ($N_H$) estimate of $N_H=9.4(\pm0.9)\times10^{20}$
cm$^{-2}$.  The cluster center is given by Shara \& Drissen (1995) as
16:17:02.48,-22:58:33.8 (J2000).  
Ferraro et al. (1999, 2003) have noted the unusually large number of
centrally concentrated blue
 stragglers in M80, which are thought to have formed through
 collisions or dynamical hardening of close binaries.  M80 is also 
unusual in having a known nova outburst (Nova 1860 T Sco; see Shara \&
Drissen 1995).  A ROSAT
 observation showed it to have at least one X-ray source in the core 
of luminosity $L_X\sim10^{32.8}$ ergs s$^{-1}$ (Hakala
et al. 1997, Verbunt 2001).  

In \S 2 we describe our observations and analysis of the globular
cluster M80.  We discuss our findings and compare them to 47 Tuc in
\S 3, and provide a summary in \S 4.

\section{M80 Observations and Analysis}
We observed M80 with \Chandra for 48.6 kiloseconds on Oct. 6, 2001 with 
the ACIS-S array at the focus for maximum soft-photon sensitivity.  We
reduced and analyzed the data using the  
\Chandra Interactive Analysis of Observations (CIAO)\footnote{Available
at http://asc.harvard.edu/ciao/.} software.  We reprocessed the level 1
event files using the latest gain files and without the pixel
randomization which is applied in standard data processing, and
filtered on grade, status, and good time intervals supplied by
standard processing.  We searched for, but did not find, times of
elevated background.  We selected an energy band of 0.5-4.5 keV to 
search for sources with maximum sensitivity while minimizing the 
background.  We ran two wavelet detection algorithms, the CIAO task {\it 
wavdetect} (Freeman et al. 2002), and the {\it pwdetect}\footnote{Available at 
http://www.astropa.unipa.it/progetti\_ricerca/PWDetect/} algorithm 
(Damiani et al. 1997), on ACIS chip 7, with similar results.  Outside
the cluster half-mass radius we select a detection sensitivity 
designed to identify a maximum of one spurious source on each chip.
In this paper we do not analyze the other four active chips.

\subsection{Detection and colors}

We analyze the sources within the half-mass radius of the cluster
(39\arcsec, Harris 1996) in 
detail, where we expect only 0.8 background AGN above 5 counts in our 
0.5-4.5 keV detection band (from Giacconi et al. 2001).
Since globular cluster X-ray   
sources are generally more massive than the typical cluster star, they
tend to concentrate towards the center of dynamically relaxed clusters.  Within
the half-mass radius we increase our  
detection algorithms' sensitivity to identify of order one spurious 
source, to calculate positions of as many cluster sources as
possible.  However, several sources obvious to the eye remained
undetected, so we increased the sensitivity to allow calculation of
source positions, and applied both algorithms in several energy
bands.  We compiled a list of robust (significance $>1.65\sigma$, more
than 3 counts, and visually confirmed) source detections to give a final 
tally of nineteen sources within the half-mass radius. These sources,
and the  extraction regions used for later analysis,
are shown in Figure 1 along with the core and half-mass regions (large
circles; astrometry is as calculated in \S 2.2 below).  We give these
sources shorthand names (e.g. CX12) which we 
will use for the rest of this paper, descending with decreasing 
counts in the 0.5-4.5 keV band.   The cluster source 
names, positions, counts in three bands, and luminosities (calculated
as below) are listed in Table
1. Note that excess unresolved emission remains in the core, probably 
representing numerous undetected sources, of which the brightest may
contribute up to 7 
counts.  At least two sources (CX9 and CX14) could be
combinations of multiple sources; however, we expect the bulk of
the counts in each to be due to a single source.  

We used extraction regions of 1\farcs25 radius circles for most sources,
except for fainter sources in the core and near brighter sources where
confusion was an issue, where we used 1\farcs or 0\farcs75 extraction
regions. We extract the counts of 
our identified sources in four bands; a soft band (0.5-1.5 keV), a 
hard band (1.5-6 keV), a detection (medium) band (0.5-4.5 keV), and
the {\it ROSAT} band (0.5-2.5 keV).  We define an
Xcolor (following GHE01a) as 2.5$\times$log(0.5-1.5 keV counts/1.5-6 
keV counts).  Our exposure map is uniform to
within 1\% between the locations of different cluster sources, so we
do not make 
exposure corrections to the observed counts.  We extracted counts from
a large source-free adjacent background region to estimate the
background flux, finding 0.019 counts/pixel in the 0.5-4.5 keV band,
0.011 counts/pixel in the 0.5-1.5 keV band, and 0.012 counts/pixel in
the 1.5-6 keV band.  Since the chance of having even one background
count recorded in any band is less than 25\% for our source
extraction regions, we do not perform background subtraction upon our
extracted numbers of counts, although we do extract background spectra
for spectral fitting purposes.  We derive aperture corrections
 from the percentage of a CXC point-spread function that falls
within our extraction circles for 1.6 keV (the mean energy of the core
sources), and apply these to the luminosities below.  

We also list the positions, colors, and exposure-corrected photon
fluxes of sources outside the half-mass radius of the cluster, but on
ACIS chip 7, in Table 2.  These are derived using the wavelet detection
program {\it pwdetect}, with the final
detection significance set to 4.5$\sigma$, leading to an expectation
of less than one false source per field.  A few spurious or 
multiply-detected sources were removed by hand, leaving 52 sources
outside the cluster half-mass radius.  
The density of 10-count sources on the rest of the chip (0.55 arcmin$^{-2}$)
indicates that 2.2 sources should be found between 1 and 2 cluster
half-mass radii, while 3 are found.  (Only 0.7 sources are expected
within the half-mass radius.)  Beyond 2 half-mass radii the
source numbers are equal to or lower than the mean chip density of
10-count sources.  The 1-2 half-mass radii overdensity 
is not significant at even the 1$\sigma$ level, but we cannot
rule out that one or two of these sources are associated with the cluster.
Assuming a power law spectrum
with photon index 1.7 and the cluster $N_H$, of order 22 background
AGN should be detected above 10 counts in our band over the entire
chip (Giacconi et al. 2001).  The 38 
sources outside the cluster half-mass radius are thus a significant
overdensity, implying a galactic population of X-ray sources in line
with the results of ongoing galactic plane and bulge surveys (e.g. Grindlay et 
al. 2003). However, further analysis of these sources is outside the
scope of this paper.
 In line with analyses of other clusters (e.g. Pooley et al. 2002a, b,
2003), we
restrict ourselves in the subsequent analysis to the sources within
the cluster half-mass radius, where massive objects will settle in
dynamically relaxed clusters such as M80.  

  We create two
versions of an ``X-ray color magnitude diagram'' to assist with source
classification.  In the first
version we follow the formalism of GHE01, assigning the logarithm of
the number of counts in the 0.5-4.5 keV band to the y-axis and 2.5
times the logarithm of the ratio of the numbers of counts in the
0.5-1.5 keV and 1.5-6 keV energy bands to the x-axis (Figure 2). This
version explicitly uses the observational quantities.  In
the second version we attempt to correct the color uniformly for the cluster
absorption.  We use the \Chandra proposal tool PIMMS
\footnote{Available at http://asc.harvard.edu/toolkit/pimms.jsp} to
investigate the effects of 
an absorbing column of $9.4\times10^{20}$ cm$^{-2}$ upon the numbers  
of detected counts in our chosen bands.  We use 0.2 and 0.3 keV
blackbody spectra, 1, 5, and 10 keV bremsstrahlung spectra, and power
law spectra with photon index 1 or 2, which cover the range of
spectral types seen in globular clusters.    
We calculate the average difference between the
calculated colors and the colors without absorption as 0.305$\pm.025$, and
calculate new corrected colors for our sources.  Instead of using counts as the
y-axis, we use the 
luminosity in the 0.5-6 keV band, where \Chandra has its greatest
sensitivity.  This provides us with an X-ray CMD which can be compared
directly to CMDs of other clusters (Figure 3).   

We compare figure 3 with the results from 47 Tuc (GHE01, Edmonds et
al. 2003a,b), $\omega$ Cen (Rutledge et al. 2002, Cool et al. 2002), NGC 6397
(GHE01b), NGC 6752 (Pooley et al. 2002a), NGC 6440
(Pooley et al. 2002b), and M28 (Becker et al. 2003).  Quiescent LMXBs
have been identified in globular clusters by their blackbody-like
spectra and high $F_X/F_{Opt}$ values.  CX2 and CX6 have similar
colors and luminosities to qLMXBs identified in 47 Tuc (X5, X7),
$\omega$ Cen (\#3), NGC 6397 (U24), and NGC 6440 (CX1) by these means,
so we classify them as probable qLMXBs.  As a further check upon our
classification, we plot in Figure 3 theoretical tracks for 
10 and 12 km nonmagnetic hydrogen-atmosphere models of Lloyd (2003).
These are essentially cooling tracks for neutron stars, 
since they show how the X-ray color of the qLMXB should change as the
luminosity decreases for a NS of fixed radius.  Clearly CX2 and CX6
are in agreement with the predictions of these tracks. 

Harder sources (-1$<$Xcolor$<$1) associated
with these clusters above $10^{32}$ ergs s$^{-1}$ seem to be almost entirely
CVs (GHE01a, GHE01b, Pooley et al. 2002a), so we identify CX2, CX3,
CX4, and CX5 as probable CVs.  Two 
eclipsing CVs in 47 Tuc (W8 and W15; GHE01, Edmonds
et al. 2003a, b) show Xcolor$<$-1, due to high intrinsic absorption of
the X-rays from the inner disk and/or WD passing through the edge-on
accretion disk.  X-ray spectra showing these colors without high
intrinsic absorption (from an accretion disk or other gas in the
system) are highly implausible.  CX15 shows 
similar colors and luminosities to these systems, so we propose it is
also a CV.  The remaining sources, below $L_X=10^{32}$ ergs s$^{-1}$,
are similar in colors and luminosity to both CV and bright AB systems in
47 Tuc, $\omega$ Cen, NGC 6397, and NGC 6752.  Soft MSPs in 47 Tuc and NGC
6752 are uniformly less X-ray luminous than any systems in M80,
but the unusual nonthermally-emitting MSPs in NGC 6397 (GHE01b) and 47
Tuc (47 Tuc-W, Edmonds et al. 2002b) are as luminous
as our faintest 
sources. (A luminous, young MSP like PSR B1821-24 in M28 should  
probably be detected in the current radio searches of M80 by N. D'Amico 
et al. We believe such objects to be rare in globular clusters, but we
cannot exclude such an object, as our 3.2 s readout time does not
allow pulsation searches at appropriate periods.)  Therefore we regard our
remaining sources as probably predominantly CVs, with some ABs and
perhaps MSPs mixed in.  We identify probable CVs, qLMXBs, and
unidentified sources with $\blacktriangle$, $\times$, and $\star$
symbols respectively in Figures 2 and 3.

\subsection{Astrometry and a possible counterpart}

  The ROSAT X-ray source \#7, 
identified by Verbunt (2001) as star HD 146457 ($V$=8.46), is clearly
detected 4\farcm2 off-axis as CXOU J161714.6-225520.
Five other serendipitous ROSAT sources also appear in the 
\Chandra field of view.  No other bright sources are unambiguously
identified with SIMBAD objects, so we use HD 146457 to define our
astrometry.  We find an offset between the \Chandra {\it wavdetect}
and {\it pwdetect} 
positions, and the Tycho Reference Catalog position, of -0.002s,
+1\farcs66 (Tycho-\Chandra), and add this offset to our nominal
astrometric solution to derive a corrected astrometric solution which
we use for the rest of this paper.  The uncertainty in the {\it
pwdetect}-derived position of HD 146457 is $\Delta\alpha$=0\fs02,
$\Delta\delta$=0\farcs3, but our 
absolute astrometric errors may be slightly increased due to
uncertainties in the plate scale and off-axis point-spread function
modeling.  Analysis of numerous point sources with optical
counterparts by Feigelson et al. (2002) and Muno et al. (2003) suggest
that typical relative astrometric uncertainties at 4\farcm off-axis are of
order 0\farcs5.  

Although classification by color and luminosity can identify some
X-ray sources with certain populations, optical identification of
counterparts is necessary to be certain of most classifications.  The
full task is beyond the scope of this work, but we do consider
previously identified possible counterparts. 
Shara and Drissen (1995) identified two faint blue stars in M80 that
are candidate CVs.  They identify one, at 
16:17:02.83, -22:58:31.3 (J2000, using the Guide Star Catalog I), as
the probable counterpart of Nova 1860
T Sco, based on a contemporary determination (Auwers 1862) of the nova
position with respect to two bright stars and the cluster center .  
Shara \& Drissen's preferred extrapolation of the Auwers (1862) nova
position (using offsets from bright stars) 
is 16:17:02.82,-22:58:32.1. These positions are respectively
1\farcs4 and 0\farcs6 away from our position for CX1, the brightest
candidate CV in our image.  Considering the uncertainties (often 1-2'') in the Guide
Star Catalog I, and in our astrometric solution above, we suggest that CX1 may
be the X-ray counterpart of Nova 1860 T Sco. Hakala et al. (1997)
provide three arguments against the identification of Nova 1860 T Sco
with the (confused) ROSAT M80 X-ray source: the positional discrepancy, the
rather high X-ray luminosity, and the rather high X-ray to optical
flux ratio.  The positional discrepancy is greatly reduced by the
resolution of the ROSAT M80 source into numerous sources by
Chandra. The X-ray luminosity of CX1 is only $3.1\times10^{32}$ ergs
s$^{-1}$ (0.5-2.5 keV), compared to the total cluster $L_X$(0.5-2.5
keV)$\sim8.6\times10^{32}$ ergs s$^{-1}$ for the ROSAT PSPC
observation cited by Hakala et al. (1995).  While high, this
luminosity is comparable to that of probable CVs
in other globular clusters, e.g., 47 Tuc (GHE01a), NGC
6440 (Pooley et al. 2002a), and Terzan 5 (Heinke et al. 2003b).
Finally, the (absorbed 0.5-2.5 keV) X-ray to (uncorrected V-band)
optical flux ratio ($F_X/F_{\rm Opt}$) of CX1, if it is the
counterpart of Nova 1860 T 
Sco, is 4.5.  While somewhat high for field systems, this is
consistent with the range of $F_X/F_{\rm Opt}$ found for CVs in 47 Tuc
by Edmonds et 
al. (2003b), of which the brightest objects may be magnetic DQ Her
systems (GHE01a).    We note that a fainter undetected CX1 counterpart
would increase the 
$F_X/F_{\rm Opt}$ ratio, and that in any case the X-ray and optical flux
measurements are not simultaneous.  Thus we conclude that the
association is plausible, 
but unproven. Further \HST analysis is in progress to look for
additional X-ray counterpart candidates and improve the \Chandra/\HST\
astrometry.

\subsection{Spectral Fitting}

For the six brightest sources associated with the cluster (over 50  
counts), we extract source (using at least
10 counts per bin) and (off-cluster) background
spectra using the CIAO script {\it psextract}, and fit the spectra in
XSPEC (Arnaud 1996)\footnote{Available at  
http://xspec.gsfc.nasa.gov.}. We correct the effective area functions
for the time-dependent low-energy quantum efficiency
degradation\footnote{See
http://cxc.harvard.edu/cal/Links/Acis/acis/Cal\_prods/qeDeg/index.html.}
We exclude bins with most photons below 0.3 keV or above 10 keV.  
We attempt to fit three models to these
spectra, all with photoelectric absorption as a free parameter
forced to be equal to or greater than the cluster value
($9.4\times10^{20}$ cm$^{-2}$).  For all analysis in this paper, we
use photoelectric absorption X-ray cross-sections of Balucinska-Church
\& McCammon (1992) in the XSPEC phabs model.   
Our models are: a thermal bremsstrahlung spectrum as
associated with CVs; a power-law model; and a hydrogen atmosphere
model (Lloyd 2003) as appropriate for qLMXBs containing thermal
neutron stars with $B<10^{10}$ G, with the radius fixed at 10 km. 
The dichotomy between harder and softer sources apparent in the X-ray
CMDs is also clear in the spectral fitting, with CX2 and CX6 showing
good fits to the hydrogen atmosphere spectral models while CX1, CX3,
CX4, and CX5 do not.  CX2 and CX6 require large values for a powerlaw
photon index ($>5$) and very small bremsstrahlung temperatures
($<0.6$ keV), which are not consistent models for any known physical
sources at these luminosities.  CX1, CX3, CX4, and CX5 give bremsstrahlung
temperatures consistent with $\sim7$ keV or more, as appropriate for
luminous CVs, particularly magnetic CVs (Eracleous, Halpern \& Patterson 1991;
Mukai 2001).  Mekal
models (Liedahl et al. 1995) give indistinguishable results, given the
low metallicity ([Fe/H]=--1.75) and high temperatures. This result
confirms our tentative classification of 
these sources as cataclysmic variables in \S 2.1.    We note
that CX6 requires a higher $N_H$ than the cluster value for any of our
models, while the other sources are consistent with the cluster value.
Heinke et al. (2003a) note enhanced $N_H$ towards X5 and X7
in 47 Tuc, presumably from gas inside or surrounding the system.  Our
preferred spectral fits to these six sources are shown in Figure 4,
and results for all three models are listed in Table 3.

For the remaining sources within the half-mass radius (except CX15,
which has a very unusual spectrum; see \S 2.1), we
extract a combined spectrum and fit this to derive the mean spectral
shape and luminosity/count ratio.  We extract a total of 235 counts,
and fit them with a thermal bremsstrahlung model of
kT$=2.3^{+.91}_{-.64}$ keV (for fixed $N_H=9.4\times10^{20}$
cm$^{-2}$), ($\chi^2_{\nu}=1.45$ for 8 dof).  A mekal fit gives very 
similar results, while fits with a power law or blackbody require very
different column densities;. The powerlaw requires 
 $N_H=26\pm9\times10^{20}$ cm$^{-2}$ with a photon index of
2.4$^{.4}_{-.3}$ ($\chi^2_{\nu}=1.8$ for 8 
dof), while the blackbody fit requires $N_H=0^{+3}_{-0}\times10^{20}$ 
cm$^{-2}$, much less than the
cluster value.  These results indicate that the fainter sources 
have lower temperatures than the bright CVs, as expected for a mix of
active binaries and (perhaps nonmagnetic) CVs, as seen in 47 Tuc
(Edmonds et al. 2003a, b).  We use the bremsstrahlung spectral fits 
to derive fluxes.  
  To calculate the luminosities of each of the fainter sources, we
multiplied their integrated luminosity by the ratio of each 
source's counts to the combined source counts (Table 1). We do this
for both the 0.5-2.5 keV band and the 0.5-6 keV band.   Derived
luminosity errors are simply Poisson or Gehrels errors from the
detected counts, without including spectral uncertainties, and are
thus underestimates.

\subsection{Time Variability}

We extracted event files from each detected source within the
half-mass radius and tested them using the IRAF {\it vartst} to
attempt to disprove the hypothesis that the source flux is constant.
Two sources (CX1 and CX8) showed variability at the 99\% 
confidence level according to both the Cramer-von Mises test and
Kolmogorov-Smirnov tests (Daniel 1990).   CX4 showed variability at
the 90\% confidence level in both tests, while no other source showed
evidence of variability.  We present the lightcurves
from these three sources, plus the (nonvariable) lightcurve from CX2
(a probable qLMXB) 
in Figure 5.   Clear flares are present in all three of the
variable sources.  X-ray flaring may be present in either CVs or
ABs, but is not expected from MSPs.  The large flare visible from CX8
is reminiscent of a flare from an AB, but we cannot make any firm
statements about these sources from their variability alone. The 
Cramer-von Mises and Kolmogorov-Smirnov tests are naturally 
far more sensitive to variability from bright sources than faint
sources, so the lack of identified variability from faint sources does
not indicate that they did not vary during the observation.

\subsection{Spatial distribution of X-ray Sources}

The radial distribution of X-ray sources in a dynamically relaxed
cluster allows an estimate of the average mass of the X-ray sources.
\citet{heinke03c} describe a procedure for estimating the typical qLMXB
mass from the spatial distribution of a sample of 20 qLMXBs in seven
clusters.  This procedure is based on maximum-likelihood fitting of a
parameterized form to the radial profile of the source distribution.
The key parameter is the ratio $q = M_X/M_\ast$ of the source mass to
the mass of the typical star that defines the optical core radius.
The approach assumes that the spatial distribution of these typical
stars is well described by a classic \citet{king66} model, which is
the case for M80 \citep{ferraro99}.  The radial profile of the source
surface density takes the form,
\begin{equation}\label{profile_fcn}
S(r) = S_0 \, \left[1 + \left({r \over r_{c\ast}}\right)^2
\right]^{(1-3q)/2},
\end{equation}
where $S_0$ is an overall normalization and $r_{c\ast}$ is the optical
core radius determined for turnoff-mass stars.  For M80,
\citet{ferraro99} have obtained $r_{c\ast}=6\farcs5$.  

In fitting the radial profile of the source distribution in M80, it is
necessary to correct the source sample for background contamination,
and ensure a uniform completeness limit.  We address the latter by
using only sources with more than 10 counts, as we are complete to
this flux limit from the cluster core out to four half-mass radii.
The expected number of background sources above 10 counts is 0.7 sources
within the half-mass radius, 2.2 between $1-2\,r_h$, 3.7 between
$2-3\,r_h$, and 5.2 between $3-4\,r_h$.  We correct for background
using the Monte-Carlo procedure described by \citet{grindlay02}.  This
procedure is carried out as part of the bootstrap resampling experiment
that is used to estimate the confidence ranges for the fit parameters.
For each of 1000 bootstrap resamplings of the source distribution, a
number of background objects is selected from a Poisson distribution
with the adopted mean value for the region under consideration.  A set
of background object positions is then generated with a uniform random
distribution over this region and the sources that are closest to these
positions are removed from the sample for that fitting trial.

Since the number of background sources beyond $2\,r_h$ is comparable
to the total number of sources detected there, we have confined our
fits to the region inside of $2\,r_h$.  The results are nearly
identical for the regions $0-1\,r_h$ and $0-2\,r_h$, with slightly
smaller errors for the former.  For this case, we obtain a mass ratio
of $q=1.44\pm0.22$ (1-$\sigma$) with a 90\% confidence range of
$1.2-2.0$.  For an assumed turnoff mass of approximately $M_\ast =
0.8\,\Msun$, the inferred source mass is $M_X = 1.2\pm0.2\,\Msun$.\@ The
90\% confidence interval extends up to 1.6\,\Msun.  For comparison,
\citet{heinke03c} find $q=1.9\pm0.2$, corresponding to $M_X =
1.5\pm0.2\,\Msun$ for the qLMXB sample.  While the difference in
inferred mass between the M80 source sample and the pure qLMXB sample
is not significant, it is in the expected direction if the former is
dominated by CVs, which should have generally lower masses than
qLMXBs.

Figure 6 shows the background-corrected cumulative
distribution of \Chandra\ sources out to $2\,r_h$, along with the
excellent fit provided by Eqn.~(\ref{profile_fcn}).  Also shown is the
analytic King model that describes the distribution of the
turnoff-mass stars.  The strong central concentration of the \Chandra\
sources, relative to the turnoff mass stars, is readily apparent.  The
source distribution is strikingly similar to the well-determined
distribution of 305 blue stragglers in M80 shown in Fig.~3 of
\citet{ferraro99}.  Thus, the masses of the \Chandra\ sources are
likely to be quite similar to those of the blue stragglers.

\subsection{Luminosity Function and Unresolved Sources}

Pooley et al. (2002b) recently showed significant differences between
the luminosity functions of several globular clusters, particularly
between those of NGC 6397 and 47 Tuc.  Following the method of Johnston \& Verbunt
(1996), they derive power-law luminosity functions $dN \propto
L_X^{-\gamma} d{\rm ln}L_X$.  Johnston \& Verbunt (1996) found
$\gamma\sim0.58$ for 14 sources in 12 globular clusters, with rather
large uncertainties, while Pooley et al. (2002) derive $\gamma$s
ranging from 0.78$^{+0.16}_{-0.17}$ for 47 Tuc to
0.29$^{+0.11}_{-0.08}$ for NGC 
6397, while NGC 6440 and NGC 6752 show intermediate values.  We use
the same method to constrain the luminosity function of M80, using a
minimum luminosity of $L_X(0.5-2.5)=1.5\times10^{31}$ ergs s$^{-1}$.  We find a
 $\gamma$ of 0.65$^{+0.30}_{-0.20}$ as our best fit (KS
probability=92\%), with values 
of $\gamma$ between 0.375 and 1.20 having KS probabilities greater than
10\%.  Using the 0.5-6 keV luminosities instead of 0.5-2.5 keV, with
 a limiting luminosity of
$L_X(0.5-6)=2.0\times10^{31}$ ergs s$^{-1}$, gives a best-fit $\gamma$ of
0.575$^{+0.23}_{-0.15}$ 
(KS probability=90\%), with an acceptable range from 0.35 to 0.975.
These limits are not greatly constraining, but suggest that M80's
overall luminosity function is less similar to that of NGC 6397 than
to the other clusters.  

We address the issue of unresolved sources in the cluster core, which
are clearly 
visible in Figure 1.  We extract a total of 48 counts in the 0.5-1.5
keV band from the core outside our source regions, and 37 counts in
the 1.5-6 keV band.  The background expected in such an area (from
measurements offset from the cluster) is 5 soft counts and 4 hard
counts.  The expected contribution from the wings of the known cluster
core sources is 23 counts in the soft band and 26 in the hard band.
This leaves a total of 20$\pm9$ soft counts and $7\pm8$ hard counts
for the remaining core sources.  (Excess emission between the core and
half-mass radii cannot be determined well due to low
statistics.) Visual inspection of images of the core in soft and hard
bands gives the impression of additional soft sources up to perhaps 6 
counts, while no undetected hard sources above 3 counts are apparent.  

Although the statistics are insufficient for firm conclusions,
these observations suggest that M80 has a population of fainter,
softer sources than the 
identified sources.  This is similar to the results from 47
Tuc presented by GHE01 and Grindlay et al. (2002).  Such faint soft
X-ray sources are likely a mixture of active binaries, MSPs, and some
CVs (Edmonds et al. 2003a, b).  We judge our
completeness limits to be roughly $L_X(0.5-2.5)=1.5\times10^{31}$ ergs
s$^{-1}$ and $L_X(0.5-6)=2.0\times10^{31}$ ergs s$^{-1}$ in the core,
with our detection and completeness limit outside the core a factor of
2 lower.  
The total 0.5-2.5 keV luminosity of our unresolved M80 core emission may
be of order $2\times10^{31}$ ergs s$^{-1}$, using a 1 keV
Raymond-Smith model in PIMMS.  We estimate that 25\% of the
core is included in our known-source extraction regions. 
Generalized King model radial distributions for objects of twice the
dominant cluster core mass (e.g. binaries and neutron stars compared
to $\sim0.7$ \Msun cluster stars) tend to distribute half these objects
inside one optical core radius (see Lugger, Cohn \& Grindlay 1995,
Grindlay et al. 2002, Verbunt 2002).  Assuming a similar distribution
for undetected 
M80 sources suggests a total luminosity of fainter sources 2.7 times
that detected, e.g. $L_X(0.5-2.5)\sim5\times10^{31}$ ergs s$^{-1}$.
The population of detected sources in 47 Tuc between $10^{30-31}$ ergs
s$^{-1}$ is some 68 sources totaling $2.1\times10^{32}$ ergs
s$^{-1}$ (GHE01), with an additional fainter unresolved emission of
$\sim7\times10^{31}$ ergs s$^{-1}$ (Grindlay et al. 2002).     
The total luminosity of detected and undetected sources in
M80 below $10^{31}$ ergs s$^{-1}$ (0.5-2.5 keV) may be $7\times10^{31}$
ergs s$^{-1}$.  Therefore, we find that M80 probably has a 
population of fainter X-ray sources perhaps 25\% as numerous as those in 47
Tuc.  

\section{Discussion}

The rates of close encounters between stars in globular clusters are
thought to scale with the square of the central density, the volume of
the core, and inversely with the velocity dispersion,
$\Gamma\propto\rho_0^2 r_c^3/\sigma$, or for a King model
$\Gamma\propto\rho_0^{1.5} r_c^2$ (Verbunt \& Hut 1987, Verbunt
2003).  According to this calculation, the production of close
encounter products in 47 Tuc should be 2.1 times larger than in M80.
This calculation does not account for the detailed dynamical
history of the cluster, including factors such as mass segregation, 
core collapse, and possible destruction of wide binaries in dense
environments. 
However, the similar central densities ($\rho_0$=4.82 and 4.87), central
concentration parameters (c$\sim2.0$), and total inferred masses
(M$\sim10^{6.1}$ \Msun and $10^6$ \Msun, Pryor \& Meylan 1993), for 47
Tuc and M80 respectively, make them a reasonable comparison.  We 
identify three differences between the two; a larger core in 47 Tuc
than M80, a substantial metallicity difference 
between 47 Tuc and M80 ([Fe/H] is -0.76 and -1.75 of solar
respectively, Harris 1996), and a
possibly high tidal destruction rate for M80 in the Galactic potential
(Dinescu et al. 1999; M80's orbit is somewhat chaotic, which makes
this prediction uncertain).  

The brighter X-ray population of M80 seems to be quite similar to that of
 47 Tuc (GHE01a); each have two qLMXBs and three CVs brighter than
 $10^{32}$ ergs/s.  This makes M80 somewhat richer than expected,
 given its smaller core.  Differences may appear in the
 fainter X-ray sources, where 
 M80 has 16 sources harder than qLMXBs above $10^{31}$
 ergs s$^{-1}$ while 47 Tuc has 24 (inside its half-mass radius, Heinke et
 al. in prep; GHE01a's smaller area of study identified 18).  47 Tuc
 may have perhaps four times as much X-ray emission from sources 
 below $10^{31}$ ergs/s, and may have a steeper luminosity function
 (see \S 2.5 and \S 2.6 above).  The fainter sources in 47 Tuc are
 a mixture of ABs, faint CVs, and MSPs (probably 30-40 above
 $L_X=10^{30}$  ergs s$^{-1}$, Edmonds et al. 2003b).  Given
 the existence of two accreting neutron stars in M80, it seems
 unlikely that M80 is much poorer in MSPs than 47 Tuc; but 
 radio timing surveys now underway may soon constrain the M80 MSP
 population.  Possible subtle differences between the clusters, if
 confirmed, may be caused by differences in metallicity or dynamical
 history, including destruction effects.   A larger group of globular 
 clusters is compared in Heinke et al. (2003c), and in Pooley et
 al. (2003), to investigate these 
 and other differences and their effects on X-ray source production.  
 
 M80 is unusual in having over 300 identified blue
 straggler stars in its central regions (Ferraro et al. 1999).  
 Ferraro and collaborators claim that stellar density alone cannot explain
 this large number (citing the much smaller number in 47 Tuc), and 
 suggest that the large blue straggler population in M80
 may be due to its dynamical state on the edge of core collapse.  At
 this stage 
 globular clusters are expected to destroy their binary populations to
 avert core collapse, possibly producing large quantities of blue
 stragglers (e.g. Fregeau et al. 2003).  However, we note that a
 similarly thorough search for blue stragglers over the central several core
 radii of 47 Tuc has not yet
 been done.  Ferraro et al. (1999)'s observations of M80 included
 roughly three core radii on the PC chip, and their searches also extended
 to the WF chips.  Ferraro et al. (2001)
 identified 43 blue stragglers in just one pointing of
 \HST PC images of 47 Tuc (which did not fully cover the core).  This
 included 36 $>$0.8 magnitudes in 
 $m_{F218W}$ above the turnoff, comparable in brightness to the 
 129 bright blue stragglers identified in M80 by Ferraro et al. (1999).
 Assuming a radial distribution of blue stragglers in 47 Tuc similar
 to that in M80 (Ferraro et al. 1999), we may expect some 130 bright
 blue stragglers in 47 
 Tuc, similar to the M80 population.  Considering the similarity in
 the X-ray source populations, this suggests that many blue
 stragglers are produced by the same mechanisms that produce X-ray
 binaries.  Ferraro et al. (2003) indeed find a good correlation
 between central density and blue straggler specific frequency for
 most globular clusters they study, although another formation route
 (probably primordial binaries) 
 seems to be required to explain the blue stragglers in low-density
 environments such as NGC 288 and the outer regions of M3.  It will be
 of interest to see if these other routes also produce X-ray sources.

\section{Conclusion}

The globular cluster M80 has a varied X-ray population similar to that
of 47 Tuc (GHE01a), including two soft sources that are probable
qLMXBs, numerous hard sources that are probable CVs (including one
probable high-inclination system with high extinction), and a sizable 
population of fainter X-ray sources.  The two bright soft sources fall
upon a calculated neutron star cooling track in an X-ray CMD and are
spectrally fit with hydrogen-atmosphere neutron star models.  The
brightest CV in the cluster may be the X-ray counterpart of the old
nova 1860 T Sco.  The radial distribution of the X-ray sources above
10 counts indicates an average system mass of $1.2\pm0.2$ \Msun, and
is similar to the distribution of blue stragglers in the cluster. This
is consistent with a mix of binaries containing neutron stars and lighter
binaries.  The overall X-ray population is slightly larger than 
expected when the cluster parameters are compared to those of 47 Tuc;
this may be connected to the cluster's unusual orbit.  The blue
straggler population in M80 may be similar to that in 47 Tuc, and we
hope that further theoretical and observational studies will probe the
connections between these different tracers of binary hardening and exchange.

\acknowledgments

C.~O.~H. acknowledges support from \Chandra grant GO2-3059A.

\begin{deluxetable}{lccccccr}
\tablewidth{7.5truein}
\tablecaption{\textbf{X-ray Sources in M80}}
\tablehead{
\colhead{\textbf{Source}} & \colhead{RA} & \colhead{Dec} &
\multicolumn{3}{c}{Counts} & \multicolumn{2}{c}{$L_X$, ergs s$^{-1}$} \\
 & (HH:MM:SS) & (DD:MM:SS) & (0.5-4.5) & (0.5-1.5) & (1.5-6) &
(0.5-6) & (0.5-2.5) 
}
\startdata

CX1 &  16:17:02.814(4) & -22:58:32.67(4) &   299  &   172   &  146  &  $6.5\times10^{32}(\pm6\%)$ & $3.4\times10^{32}(\pm6\%)$ \\
CX2  &  16:17:02.576(2) & -22:58:36.48(3) &  209  &   195   &  14   &  $2.9\times10^{32}(\pm7\%)$  & $2.9\times10^{32}(\pm7\%)$  \\
CX3  &  16:17:01.597(3) & -22:58:27.95(4) &  148  &   86    &  68   &  $3.3\times10^{32}(\pm8\%)$  &  $1.9\times10^{32}(\pm9\%)$  \\
CX4  &  16:17:02.005(4) & -22:58:33.03(5) &  147  &   76    &  81   &  $3.8\times10^{32}(\pm8\%)$  & $2.0\times10^{32}(\pm9\%)$  \\
CX5  &  16:17:01.708(4) & -22:58:15.34(6) &  80   &   37    &  46   &  $2.0\times10^{32}(\pm11\%)$  & $9.5\times10^{31}(\pm13\%)$  \\
CX6  &  16:17:03.569(5) & -22:58:25.30(6) &  55   &   54    &  2    &  $9.1\times10^{31}(\pm13\%)$  & $9.1\times10^{31}(\pm13\%)$   \\
CX7  &  16:17:02.164(6) & -22:58:37.27(5) &  50   &   25    &  25   &  $1.0\times10^{32}(\pm14\%)$  & $5.7\times10^{31}(\pm15\%)$   \\
CX8  &  16:17:01.114(5) & -22:58:29.33(8) &  25   &   11    &  14   &  $5.1\times10^{31}(\pm20\%)$  & $3.2\times10^{31}(\pm20\%)$   \\
CX9  &  16:17:02.401(7) & -22:58:32.6(1) &   24   &   14    &  10   &  $5.0\times10^{31}(\pm20\%)$  & $2.8\times10^{31}(\pm22\%)$   \\
CX10 &  16:17:00.407(8) & -22:58:28.87(7) &  23   &   10    &  13   &  $4.7\times10^{31}(\pm21\%)$  & $2.4\times10^{31}(\pm24\%)$   \\
CX11 &  16:17:02.472(8) & -22:58:37.86(8) &  21   &   13    &  9   &  $4.6\times10^{31}(\pm21\%)$  & $2.2\times10^{31}(\pm25\%)$   \\
CX12 &  16:17:02.565(7) & -22:58:45.0(1) &  20   &   14    &  6    &  $4.1\times10^{31}(\pm22\%)$  &  $2.7\times10^{31}(\pm22\%)$   \\
CX13 &  16:17:01.755(7) & -22:58:29.29(9) &  12   &   5     &  8    &  $2.7\times10^{31}(\pm28\%)$  & $1.5\times10^{31}(\pm30\%)$   \\
CX14 &  16:17:02.553(8) & -22:58:30.5(2) &  11   &   7    &  4    &  $2.3\times10^{31}(\pm30\%)$  &  $1.5\times10^{31}(\pm30\%)$  \\
CX15 &  16:17:02.100(7) & -22:58:31.8(1) &   9   &   1     &  13   &  $3.5\times10^{31}(\pm27\%)$* & $4\times10^{30}(^{+132}_{-65}\%)$*  \\
CX16 &  16:17:02.119(8) &  -22:58:19.8(2) &  9    &   4     &  5    &  $1.8\times10^{31}(^{+46}_{-33}\%)$  & $1.1\times10^{31}(^{+49}_{-35}\%)$   \\
CX17 &  16:17:02.220(7) & -22:58:33.7(2) &   8    &   7     &  1    &  $1.7\times10^{31}(^{+49}_{-35}\%)$  & $1.1\times10^{31}(^{+49}_{-35}\%)$   \\
CX18 &  16:17:02.820(7) & -22:58:36.0(2) &   6    &   4     &  2    &  $1.3\times10^{31}(^{+60}_{-40}\%)$  & $8\times10^{30}(^{+60}_{-40}\%)$   \\
CX19 &  16:17:03.85(1) & -22:58:47.1(2) &    5    &   4     &  1    &  $1.0\times10^{31}(^{+68}_{-43}\%)$  & $7\times10^{30}(^{+68}_{-43}\%)$   \\
								 	
\enddata
\tablecomments{
Names, positions, counts in three X-ray energy bands (energies given
in keV),
and estimated luminosities of X-ray sources within the half-mass
radius of M80.  The errors in parentheses after the position represent
the $1\sigma$ uncertainties in the relative positions of the sources,
derived from {\it wavdetect} results.  The counts in each band are the
numbers of photons within the circular source regions of Figure
1. Luminosities have been adjusted to account for the percentage of
the point spread function included in each region. The
luminosities for CX1-CX6 are derived from individual spectral fitting,
while the luminosities for CX7-CX19 are derived from fitting their 
combined (except CX15) spectrum.  Luminosity errors (given in percentage) are
derived from Poisson or Gehrels statistics of the detected counts in each band.  
*-CX15 probably suffers significant intrinsic absorption, unaccounted
for in these luminosities (see text).
}
\end{deluxetable}

\begin{deluxetable}{lcccr}
\tablewidth{5.5truein}
\tablecaption{\textbf{Serendipitous Sources in the M80 Field}}
\tablehead{
\colhead{\textbf{Name}} & \colhead{RA}  & \colhead{Dec}   &
\colhead{Counts} & \colhead{Flux}  \\
(CXOU J) & (HH:MM:SS) & (DD:MM:SS) & (0.5-4.5 keV) & (phot 
cm$^{-2}$ s$^{-1}$) \\
}
\startdata
 161646.9-225737  &  16:16:46.93(2)   & -22:57:37.7(3)   &	 426.3(22.8) & 1466.6(78.5) \\
 161646.9-225509  &  16:16:46.98(12)  & -22:55: 9.9(18)  &	  12.5(5.0)  &  42.6(16.9)  \\
 161647.1-225535  &  16:16:47.19(7)   & -22:55:35.9(10)  &	  83.2(9.2)  & 280.2(30.8)  \\
 161648.3-225906  &  16:16:48.35(7)   & -22:59: 6.2(10)  &	   7.6(3.6)  &  25.8(12.3)  \\
 161648.5-225311  &  16:16:48.55(14)  & -22:53:11.9(21)  &	  24.7(7.8)  &  95.2(30.1)  \\
 161648.5-225752  &  16:16:48.55(5)   & -22:57:52.4(8)   &	  17.7(5.9)  &  60.4(20.2)  \\
 161648.6-225412  &  16:16:48.64(11)  & -22:54:12.4(17)  &	  17.0(6.1)  &  61.8(22.3)  \\
 161649.8-225705  &  16:16:49.86(2)   & -22:57: 5.3(3)   &	   4.2(3.2)  &  14.4(11.0)  \\
 161650.2-225349  &  16:16:50.28(9)   & -22:53:49.5(14)  &	  10.7(4.8)  &  37.0(16.7)  \\
 161652.1-225612  &  16:16:52.15(2)   & -22:56:12.9(3)   &	   4.2(3.7)  &  14.0(12.4)  \\
 161652.2-225614  &  16:16:52.29(3)   & -22:56:14.6(5)   &	 104.3(10.7) &  350.5(36.1) \\
 161652.8-225420  &  16:16:52.85(12)  & -22:54:20.8(18)  &	  11.8(4.7)  &  42.1(16.9)  \\
 161653.8-225844  &  16:16:53.83(2)   & -22:58:44.2(4)   &	  23.1(7.0)  &  80.3(24.2)  \\
 161653.9-225618  &  16:16:53.93(5)   & -22:56:18.8(8)   &	  18.5(6.4)  &  62.3(21.5)  \\
 161653.9-225904  &  16:16:53.94(2)   & -22:59: 4.1(3)   &	  27.2(5.1)  &  92.2(17.3)  \\
 161654.6-225615  &  16:16:54.63(6)   & -22:56:15.2(10)  &	   8.5(3.6)  &  28.4(12.2)  \\
 161655.0-225451  &  16:16:55.09(6)   & -22:54:51.6(9)   &	 118.4(11.5) &  398.0(38.6) \\
 161655.5-225925  &  16:16:55.58(1)   & -22:59:25.7(2)   &	  69.5(8.9)  & 235.5(30.1)  \\
 161655.8-225625  &  16:16:55.85(3)   & -22:56:25.1(4)   &	   5.0(2.8)  &  16.6(9.4)  \\
 161655.9-225635  &  16:16:55.97(3)   & -22:56:35.8(4)   &	  11.7(4.6)  &  41.6(16.3)  \\
 161658.3-225838  &  16:16:58.30(3)   & -22:58:38.1(4)   &	   5.6(3.0)  &  18.6(9.9)  \\
 161659.1-225349  &  16:16:59.14(17)  & -22:53:49.5(25)  &	  15.0(6.0)  &  52.9(20.9)  \\
 161659.8-225931  &  16:16:59.88(2)   & -22:59:31.1(2)   &	  10.0(4.3)  &  32.9(14.1)  \\
 161700.8-225700  &  16:17: 0.90(5)   & -22:57: 0.1(7)   &	   7.4(3.3)  &  29.1(13.0)  \\
 161701.0-225307  &  16:17: 1.05(16)  & -22:53: 7.8(24)  &	  16.2(6.0)  &  57.9(21.5)  \\
 161702.0-230033  &  16:17: 2.05(1)   & -23: 0:33.0(2)   &	  65.7(8.9)  & 218.0(29.6)  \\
 161704.1-225527  &  16:17: 4.10(3)   & -22:55:27.4(5)   &	  86.7(11.2) &  283.0(36.6) \\
 161704.5-230055  &  16:17: 4.57(4)   & -23: 0:55.2(6)   &	   5.0(2.7)  &  16.7(9.0)  \\
 161704.7-225622  &  16:17: 4.78(4)   & -22:56:22.6(5)   &	   5.8(2.9)  &  18.8(9.4)  \\
 161705.1-225805  &  16:17: 5.10(2)   & -22:58: 5.5(4)   &	  20.9(6.6)  &  70.1(22.1)  \\
 161706.2-225835  &  16:17: 6.22(2)   & -22:58:35.6(3)   &	   8.7(4.2)  &  29.3(14.0)  \\
 161706.4-225318  &  16:17: 6.49(16)  & -22:53:18.0(24)  &	  15.2(5.6)  &  53.7(19.8)  \\
 161706.8-225808  &  16:17: 6.87(2)   & -22:58: 8.2(3)   &	  29.6(5.6)  &  99.1(18.6)  \\
 161707.0-230121  &  16:17: 7.06(3)   & -23: 1:21.3(5)   &	  24.3(7.3)  & 122.7(36.8)  \\
 161707.7-225755  &  16:17: 7.77(2)   & -22:57:55.8(3)   &	  77.6(9.3)  & 258.2(31.1)  \\
 161707.8-225752  &  16:17: 7.90(3)   & -22:57:52.0(4)   &	   4.6(2.7)  &  15.3(8.9)  \\
 161708.2-225617  &  16:17: 8.27(3)   & -22:56:17.1(4)   &	   4.5(2.6)  &  14.7(8.5)  \\
 161709.1-225529  &  16:17: 9.20(3)   & -22:55:29.1(4)   &	 247.5(19.0) &  817.7(62.7) \\
 161709.8-230034  &  16:17: 9.81(3)   & -23: 0:34.7(4)   &	  11.1(4.4)  &  36.7(14.4)  \\
 161712.3-225313  &  16:17:12.34(8)   & -22:53:13.2(11)  &	 202.9(15.5) & 1039.9(79.3) \\
 161712.5-230034  &  16:17:12.51(4)   & -23: 0:34.4(6)   &	  48.2(7.5)  & 160.6(25.1)  \\
 161713.6-225324  &  16:17:13.63(16)  & -22:53:24.1(24)  &	  18.1(6.6)  &  67.4(24.5)  \\
 161713.7-225549  &  16:17:13.79(6)   & -22:55:49.8(9)   &	  47.2(7.5)  & 156.2(25.0)  \\
 161714.6-225520  &  16:17:14.66(2)   & -22:55:20.0(3)   &    13492.3(121.7) & 47046.0(424.5) \\
 161715.8-225516  &  16:17:15.84(5)   & -22:55:16.5(7)   &	 356.0(20.5) & 1263.8(72.7) \\
 161716.0-225857  &  16:17:16.04(2)   & -22:58:57.4(2)   &	  10.4(4.4)  &  34.9(14.7)  \\
 161716.0-225858  &  16:17:16.05(2)   & -22:58:58.9(3)   &	   5.1(3.3)  &  17.1(11.0)  \\
 161716.8-225608  &  16:17:16.88(5)   & -22:56: 8.7(8)   &	   6.5(4.0)  &  21.5(13.3)  \\
 161716.9-225953  &  16:17:16.99(3)   & -22:59:53.1(5)   &	 116.4(11.9) &  413.3(42.1) \\
 161718.7-225512  &  16:17:18.71(7)   & -22:55:12.0(11)  &	  67.8(9.9)  & 229.8(33.4)  \\
 161719.5-225705  &  16:17:19.56(7)   & -22:57: 5.3(10)  &	   8.7(3.9)  &  29.0(13.0)  \\
 161721.7-225415  &  16:17:21.75(15)  & -22:54:15.2(23)  &	  19.1(6.6)  &  66.0(22.7) \\
\enddata		    			     		    
\tablecomments{Sources outside the M80 half-mass radius detected on the S3 chip during
the M80 observation.  Relative positional errors are given in parentheses on
the last quoted digits.  Counts in the 0.5-4.5 keV band,  photon
flux, and errors in both are given by {\it pwdetect} tool.  
}
\end{deluxetable}

\begin{deluxetable}{lccccccccr}
\tablewidth{7.5truein}
\tablecaption{\textbf{Spectral Fits to Brighter M80 Sources}}
\tablehead{
\colhead{\textbf{Source}} & \multicolumn{3}{c}{H-atmosphere}  &
\multicolumn{3}{c}{Bremsstrahlung} & 
 \multicolumn{3}{c}{Powerlaw}  \\
 & (kT, eV) & ($N_H\times10^{20}$) & ($\chi^2_{\nu}$/dof) & (kT, keV)
 & ($N_H\times10^{20}$) & ($\chi^2_{\nu}$/dof) 
& ($\alpha$) & ($N_H\times10^{20}$) & ($\chi^2_{\nu}$/dof) 
}
\startdata
CX1 & 106 & 39.7 & 3.97/29 & $>7.3$ & 9.4$^{+3.3}_{-0}$ & 0.80/28 & $1.4^{+.3}_{-.1}$ & $9.4^{+5.6}_{-0}$ & 0.76/28 \\
CX2 & $89\pm2$  & $9.4^{+2.5}_{-0}$ & 0.56/18  &  $0.43^{+.11}_{-.13}$
& $11.5^{+14.1}_{-2.1}$ & 0.55/17 & $6.3^{+1.1}_{-1.4}$ &
$54^{+19}_{-14}$ & 0.80/17 \\ 
CX3 & 99  & 52.3  & 3.3/13 &  $6.0^{+12}_{-3.0}$ & $9.4^{+9.1}_{-0}$ & 0.43/12 &
$1.7^{+.5}_{-.3}$ & $9.4^{+15}_{-0}$ & 0.32/12 \\
CX4 & 104 & 72 & 3.88/13 &  $7.6^{+36}_{-3.9}$ & $17^{+17}_{-8}$ & 0.50/12 &
$1.7^{+.5}_{-.4}$ & $24^{+20}_{-15}$ & 0.53/12 \\
CX5 & 108  & 138  & 4.9/7 &  $>4.1$ & $12^{+19}_{-3}$ & 0.53/6 &
$1.4^{+.6}_{-.4}$ & $13^{+15}_{-4}$ & 0.57/6 \\
CX6 & $76^{+5}_{-4}$  & $22^{+8}_{-7}$ & 0.37/4 &
$0.37^{+.15}_{-.21}$ & $27^{+26}_{-16}$ & 0.53/3 & $6.2^{+4}_{-2.2}$ &
$63^{+59}_{-39}$ & 0.57/3 \\ 
\enddata
\tablecomments{Spectral fits to cluster sources, with background
subtraction, in XSPEC.  Errors are 90\% confidence for a single
parameter; spectra are binned 
with 10 counts/bin.  All fits include photoelectric absorption
forced to be $\geq9.4\times10^{20}$ cm$^{-2}$, the cluster $N_H$
derived from optical studies.  Hydrogen
atmosphere fits are made with radius fixed to 10 km. 
}
\end{deluxetable}

\clearpage


\psfig{file=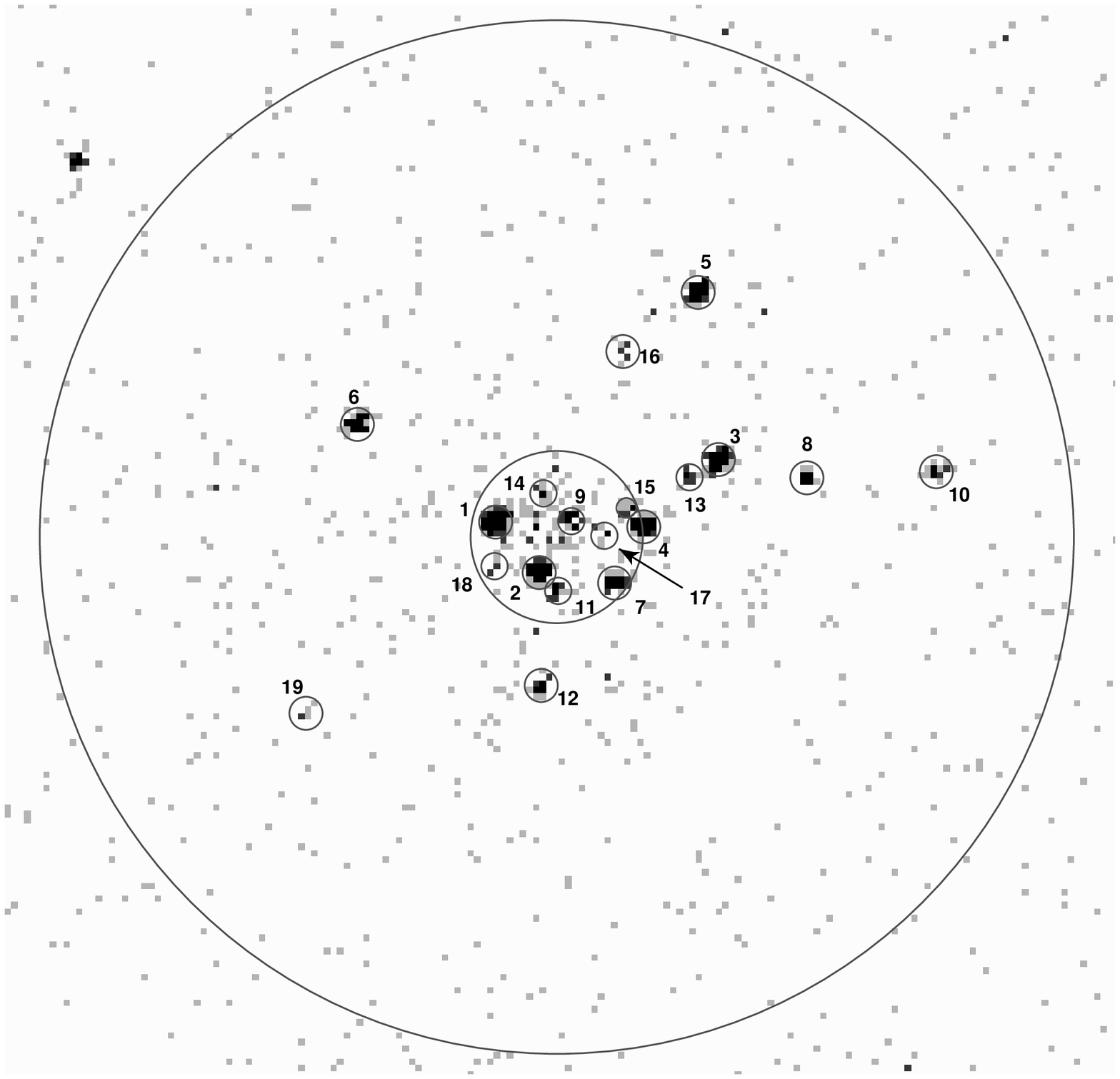,height=7in}

\figcaption[f1.eps]{\Chandra\ ACIS-S image of the globular cluster
M80.  The two larger circles represent the core (inner) and half-mass radii
of the cluster.  The 19 sources within the half-mass radius are
labeled (in order of decreasing counts in the 0.5-4.5 keV band), and
the extraction regions are overlaid.  Additional X-ray 
emission is visible from the central cluster core from sources unresolved with
{\it WAVDETECT}.  CXOU J161705.1--225805 is also visible at upper
left and may be associated with the cluster. 
\label{Figure 1}}

\psfig{file=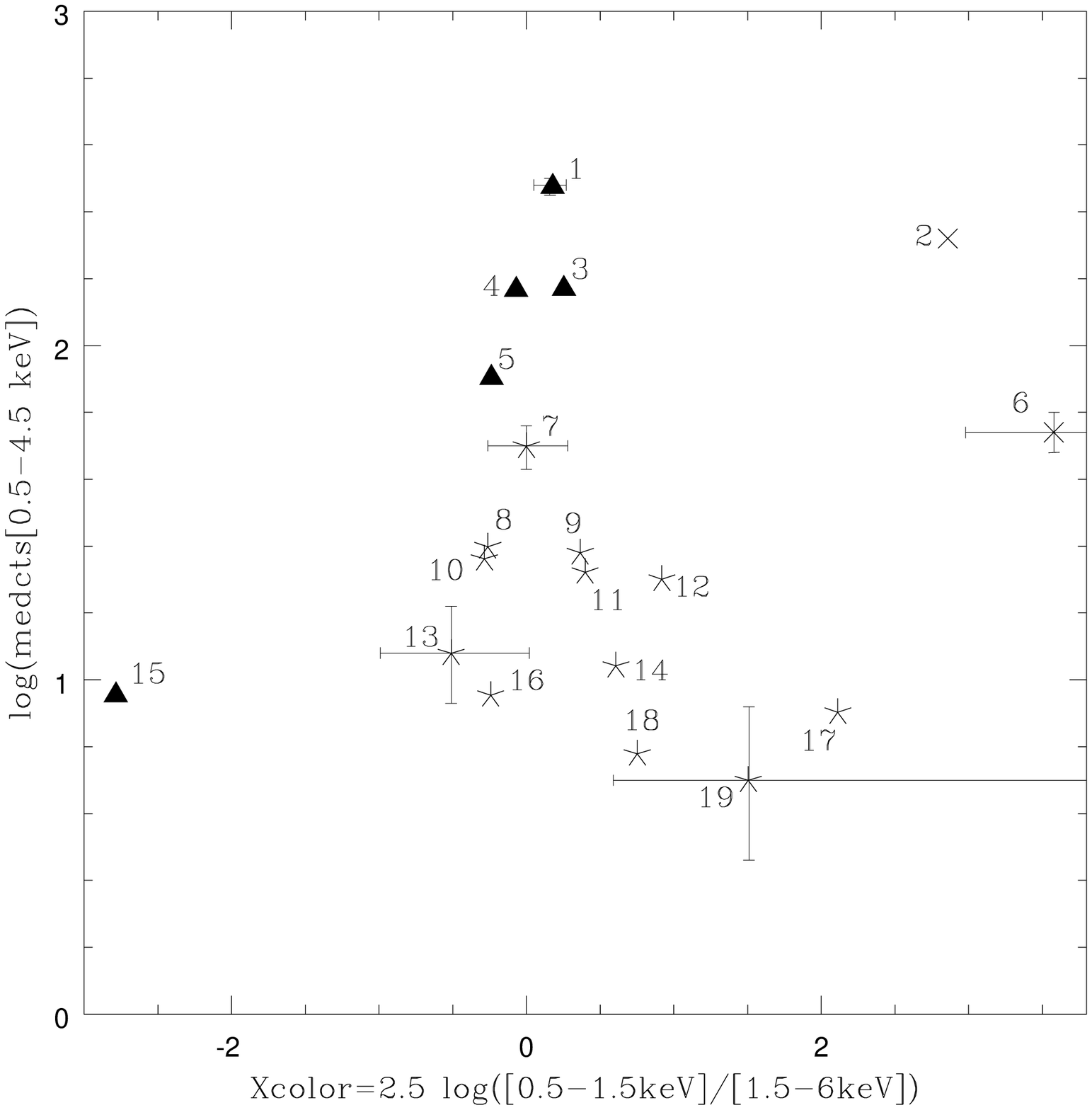,height=6in}

\figcaption[f2.eps]{Instrumental X-ray CMD for the 19 sources in
the globular cluster M80.  Vertical axis is the log of the number of
counts detected in the 0.5-4.5 keV band, while the horizontal axis is
Xcolor, defined (following GHE01) as 2.5 times the log of the ratio of
counts detected in the 0.5-1.5 keV band over the counts detected in
the 1.5-6 keV band. A few error bars are shown, representing 1$\sigma$
errors of Gehrels (1986).  Symbols represent probable source nature, as in
GHE01; $\times$: qLMXBs, $\blacktriangle$: CVs, $\star$: ambiguous
(probably a mixture of CVs and ABs).  Sources are numbered as in Table
1.
\label{Figure 2}}

\psfig{file=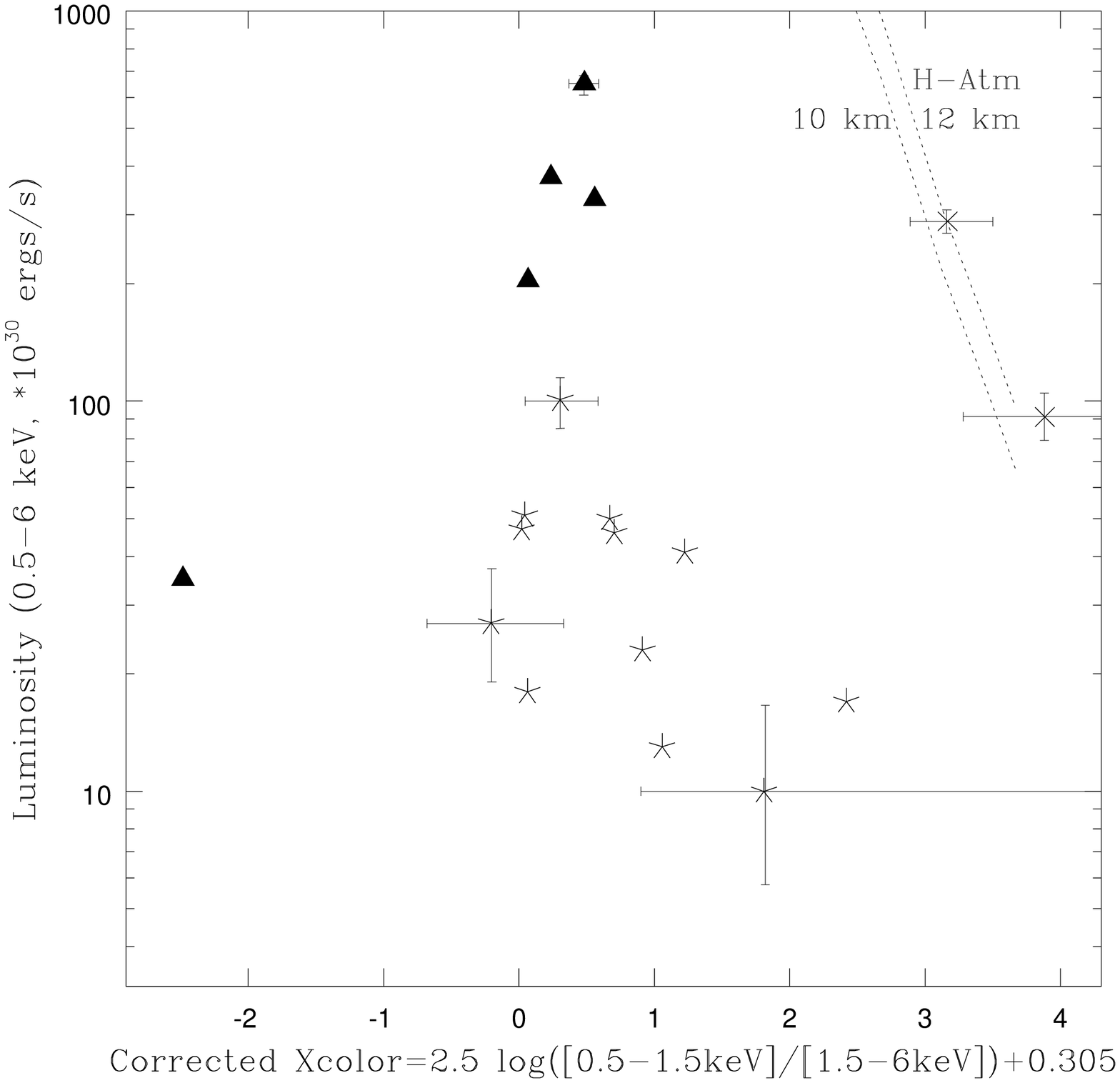,height=6in}

\figcaption[f3.eps]{Standardized X-ray CMD for the 19 sources in
the globular cluster M80.  Vertical axis is the 0.5-6 keV X-ray
luminosity in units of $10^{30}$ ergs s$^{-1}$, derived from spectral
fitting (see text).  Horizontal axis is Xcolor as in figure 2, but
with a uniform shift of 0.305 added to correct for the effects of
photoelectric absorption (not including intrinsic absorption).
Errors do not include spectral uncertainties.  Symbols as in Figure 2.  
\label{Figure 3}}

\psfig{file=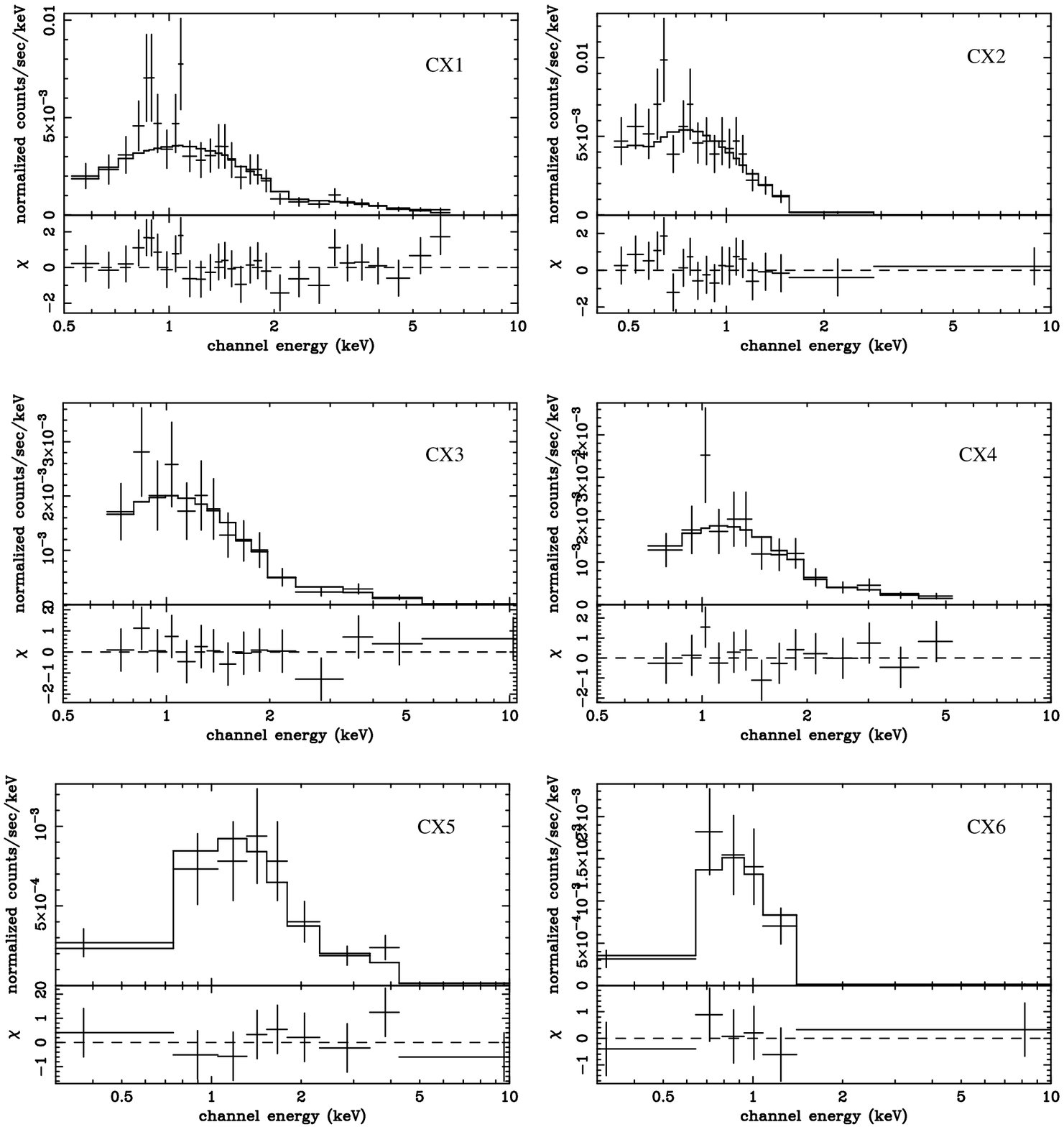,height=7in}

\figcaption[f4.eps]{Energy spectra of six of the brighter sources in M80.
Upper panels show the data compared with a nonmagnetic hydrogen
atmosphere neutron 
star model (Lloyd 2003) for CX2 and CX6, and a thermal bremsstrahlung
model for CX1, CX3, CX4 and CX5.  Lower panels show the contributions
to the $\chi^2$ statistic for each fit.  Photoelectric absorption is
included in each fit.
\label{Figure 4}}

\psfig{file=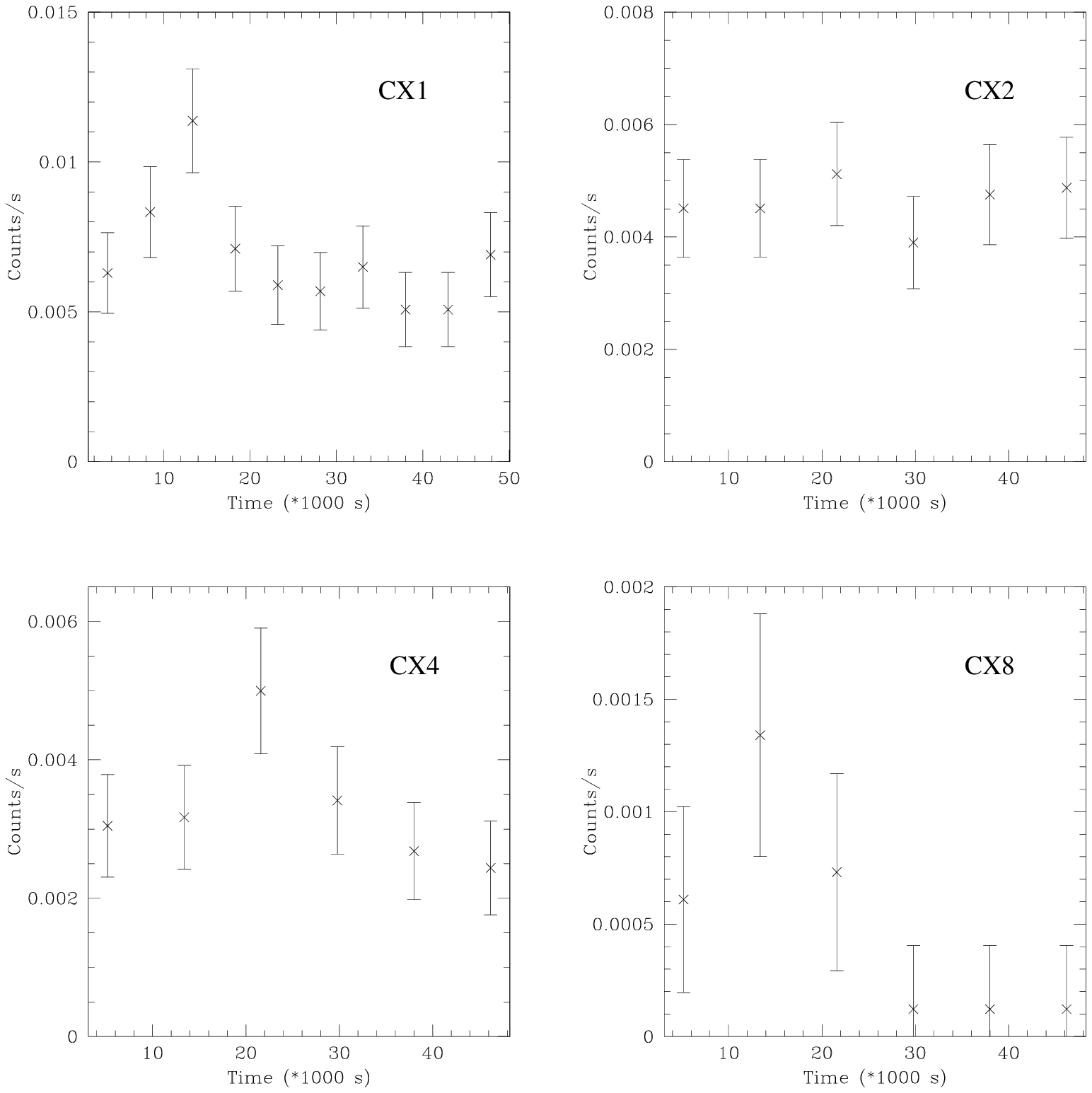}

\figcaption[f5.eps]{Lightcurves for three variable sources in
M80, plus the probable qLMXB CX2 for comparison.  CX1 and CX8 are
found to be variable at or above the 99\% level in two variability
tests, while CX4 appears to be variable at the 90\% level in both
tests.  All three of the variable sources appear to show flares.  
\label{Figure 5}}

\begin{figure}
\plotone{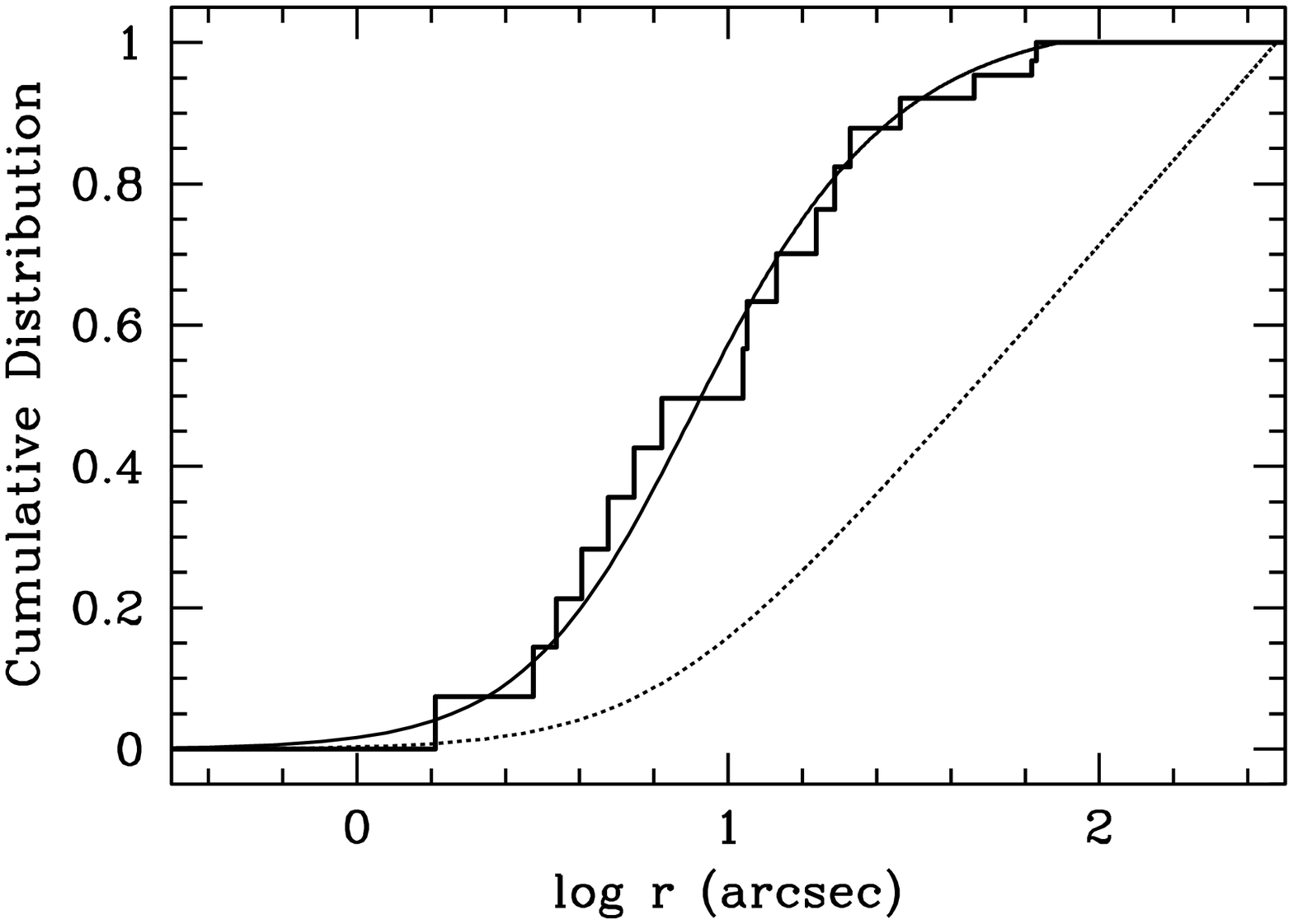}
\label{m80_profile}
\caption{Profile fit to the background-corrected M80 source
distribution.  The histogram is the average of 1000
background-corrected resamplings of the original source
distribution. The smooth solid line is the fit of
Eqn.~(\ref{profile_fcn}) with $q=1.44$.  The dotted line shows the
distribution of the turnoff-mass stars, i.e.\ Eqn.~(\ref{profile_fcn})
with $q=1$.  The latter curve is normalized so that the sample is
complete at $300''$, the approximate outer limit of the
\citet{ferraro99} profile.}
\end{figure}

\end{document}